\documentclass[aps,pra,twocolumn,showpacs,superscriptaddress,longbibliography]{revtex4-1}
\usepackage{graphicx}    
\usepackage{dcolumn}    
\usepackage{bm}             
\usepackage{amssymb}   
\usepackage{amsmath}    
\usepackage{subfigure}    
\usepackage{braket}
\usepackage{mathrsfs}
\usepackage[export]{adjustbox}
\usepackage[toc,page]{appendix}
\begin{document}

\title{Laser-free trapped-ion entangling gates with simultaneous insensitivity to qubit and motional decoherence}

\author{R. T. Sutherland}
\email{sutherland11@llnl.gov}
\affiliation{Physics Division, Physical and Life Sciences, Lawrence Livermore National Laboratory, Livermore, California 94550, USA}
\author{R. Srinivas}
\affiliation{Time and Frequency Division, National Institute of Standards and Technology, 325 Broadway, Boulder, Colorado 80305, USA}
\affiliation{Department of Physics, University of Colorado, Boulder, Colorado 80309, USA}
\author{S. C. Burd}
\affiliation{Time and Frequency Division, National Institute of Standards and Technology, 325 Broadway, Boulder, Colorado 80305, USA}
\affiliation{Department of Physics, University of Colorado, Boulder, Colorado 80309, USA}
\author{H. M. Knaack}
\affiliation{Time and Frequency Division, National Institute of Standards and Technology, 325 Broadway, Boulder, Colorado 80305, USA}
\affiliation{Department of Physics, University of Colorado, Boulder, Colorado 80309, USA}
\author{A. C. Wilson}
\affiliation{Time and Frequency Division, National Institute of Standards and Technology, 325 Broadway, Boulder, Colorado 80305, USA}
\author{\break D. J. Wineland}
\affiliation{Time and Frequency Division, National Institute of Standards and Technology, 325 Broadway, Boulder, Colorado 80305, USA}
\affiliation{Department of Physics, University of Colorado, Boulder, Colorado 80309, USA}
\affiliation{Department of Physics, University of Oregon, Eugene, Oregon 97403, USA}
\author{D. Leibfried}
\affiliation{Time and Frequency Division, National Institute of Standards and Technology, 325 Broadway, Boulder, Colorado 80305, USA}
\author{D. T. C. Allcock}
\affiliation{Time and Frequency Division, National Institute of Standards and Technology, 325 Broadway, Boulder, Colorado 80305, USA}
\affiliation{Department of Physics, University of Colorado, Boulder, Colorado 80309, USA}
\affiliation{Department of Physics, University of Oregon, Eugene, Oregon 97403, USA}
\author{D. H. Slichter}
\affiliation{Time and Frequency Division, National Institute of Standards and Technology, 325 Broadway, Boulder, Colorado 80305, USA}
\author{S. B. Libby}
\affiliation{Physics Division, Physical and Life Sciences, Lawrence Livermore National Laboratory, Livermore, California 94550, USA}

\date{\today}

\begin{abstract}
The dominant error sources for state-of-the-art laser-free trapped-ion entangling gates are decoherence of the qubit state and the ion motion. The effect of these decoherence mechanisms can be suppressed with additional control fields, or with techniques that have the disadvantage of reducing gate speed. Here, we propose using a near-motional-frequency magnetic field gradient to implement a laser-free gate that is simultaneously resilient to both types of decoherence, does not require additional control fields, and has a relatively small cost in gate speed.
\end{abstract}
\pacs{}
\maketitle

\section{Introduction}
Trapped ions are a promising platform for quantum simulations and universal quantum computation due to their long coherence times, inherent uniformity, and high gate fidelities \cite{cirac_1995, monroe_1995,nielsen_2010, haffner_2008,blatt_2008,harty_2014,ballance_2016,gaebler_2016}. The most common method for performing high-fidelity multi-qubit entangling gates, a requirement for universal quantum processors, relies on coupling the internal qubit ``spin'' states to collective motional degrees of freedom \cite{cirac_1995, monroe_1995,wineland_1998}. Geometric phase gates\textemdash which create entanglement through closed, spin-dependent trajectories in motional phase space\textemdash are widely used because they are first-order insensitive to ion temperature (in the Lamb-Dicke limit) \cite{molmer_1999,molmer_2000,leibfried_2003}. Geometric phase gates employing laser beams to create the required spin-motion coupling have been used to generate Bell states with fidelities $\sim 0.999$ \cite{ballance_2016, gaebler_2016}, with the main error arising from off-resonant photon scattering \cite{ozeri_2007}. Alternative laser-free schemes induce spin-motion coupling with static \cite{mintert_2001,khromova_2012,lake_2015,hensinger_2015,weidt_2016,webb_2018}, near-qubit-frequency \cite{ospelkaus_2008,ospelkaus_2011,harty_2016,wunderlich_2017,hahn_2019, zarantonello_2019}, or near-motional-frequency \cite{chiaverini_2008, ospelkaus_2008,sutherland_2019,srinivas_2018} magnetic field gradients. While laser-free schemes eliminate photon scattering errors and do not require stable, high-power lasers, they can be more susceptible to other noise sources due to their typically longer gate durations.

Qubit frequency shifts or miscalibrations due to fluctuating field amplitudes are the primary sources of error in laser-free gates implemented with microwave field gradients \cite{ospelkaus_2011,webb_2018}. Recent work has shown that some of these shifts may be reduced passively through careful trap design \cite{hahn_2019}. They can also be reduced actively by adding control fields to perform dynamical decoupling \cite{viola_1998_pra,viola_1999_prl,timoney_2011,bermudez_2012,weidt_2016}; to date, the best experimentally demonstrated Bell state fidelity using such a scheme is 0.997(1) \cite{harty_2016}. Schemes which employ static magnetic field gradients to perform laser-free gates allow high-fidelity individual addressing of the ions in frequency space \cite{johanning_2009}; however, this means that the microwave frequencies needed to realize an entangling gate are different for each ion, increasing the total number of drive tones necessary to implement such a gate~\cite{weidt_2016, webb_2018}. 

Since laser-free geometric phase gates are typically slower than laser-based gates by an order of magnitude, the qubits spend more time entangled with the motional mode, and thus the gates are more sensitive to motional decoherence. Typically, after qubit frequency shifts, this is the next most important source of gate error \cite{harty_2016,weidt_2016,hahn_2019}. In the work reported here, we consider motional decoherence in three distinct regimes, depending on the timescale and nature of the decoherence: secular frequency shifts, motional heating, and motional dephasing. 

In their original proposal, S\o renson and M\o lmer pointed out that dividing a geometric phase gate into $K$ loops decreases gate errors from heating and motional dephasing \cite{molmer_2000}. Separately, decoherence from secular frequency shifts can be suppressed with Walsh sequences \cite{hayes_2012}, or with phase modulation of gate fields \cite{green_2015,leung_2018}. These techniques increase the gate duration $t_{G}$ in exchange for robustness. Polychromatic gates \cite{haddadfarshi_2016}, geometric phase gates comprised of multiple simultaneously  applied gate fields with optimized amplitudes, reduce the impact of all three types of motional decoherence and have a comparatively smaller increase in $t_{G}$. This technique was recently demonstrated for both laser-based \cite{shapira_2018} and laser-free gates \cite{webb_2018}. These gate implementations, however, remain sensitive to qubit frequency offsets, although they provide some dynamical decoupling for zero-mean fluctuations; their physical implementation also requires additional control fields, increasing experimental complexity. Similar robustness to motional errors can alternatively be achieved by smooth ramping of the gate field amplitude over the entire duration of the gate pulse \cite{zarantonello_2019}. 

\begin{figure}[tb]
\includegraphics[width=0.45\textwidth]{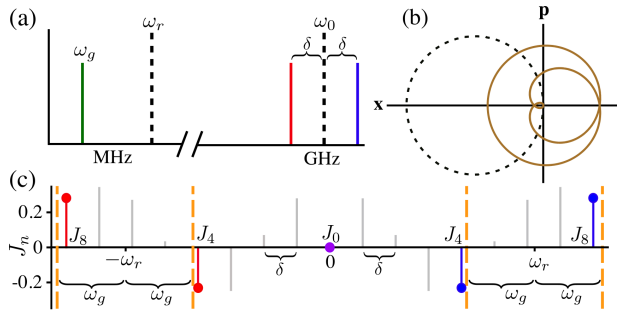}
\centering
\caption{(a) Frequency spectrum of the three fields used to perform the described gates. (b) Motional phase space trajectories of two gates that are insensitive to qubit frequency shifts. The black dashed line represents a conventional single-tone gate \cite{leibfried_2003} and the brown solid line corresponds to a motionally robust gate proposed in this work. (c) For the arguments described in the text, amplitudes of Bessel functions $J_{n}$ versus frequency in the rotating frame of the qubit. The dynamics of the bichromatic microwave pair gives an effective modulation of the microwave Rabi frequency, and causes a series of resonances at integer multiples of $\delta$ with strength proportional to the corresponding Bessel function. A polychromatic gate occurs when two resonances (here $J_4$ and $J_8$) are tuned near $|\omega_{r}\pm\omega_{g}|$.}
\label{fig:phase_space}
\end{figure}

In this work, we build upon Ref.~\cite{sutherland_2019} to propose a gate that provides simultaneous robustness to qubit frequency shifts and to motional decoherence without requiring additional control fields, offering a combination of increased fidelity and decreased experimental overhead. The protection from qubit frequency errors extends to much larger frequency offsets than can typically be protected against using dynamical decoupling techniques, because the entangling gate commutes with these fluctuations, unlike all other laser-free entangling gates demonstrated previously.

The outline of the paper is as follows.  In Sec.~\ref{sec:bip} we present a theoretical framework we use to analyze the qubit dynamics generated by a pair of microwave fields symmetrically detuned around the qubit frequency and a near-motional-frequency magnetic field gradient. In Sec.~\ref{sec:poly_1}, we show how to tune these control fields to realize a geometric phase gate which is insensitive to qubit frequency shifts and robust to motional decoherence. We also characterize its performance with numerical simulations. In Sec.~\ref{sec:poly_2}, we analyze polychromatic gates that use multiple pairs of microwave fields and contrast them with the scheme presented here. Finally, we present conclusions and prospects for future work.

\section{Bichromatic interaction picture for laser-free gates}\label{sec:bip}
We consider a Hamiltonian for laser-free gates between two trapped ions with identical qubit frequencies $\omega_{0}$, coupled to a shared motional mode with frequency $\omega_{r}$.  We apply a pair of microwave-frequency magnetic fields (symmetrically detuned around $\omega_0$ by $\pm\delta$), and a separate magnetic field gradient oscillating at $\omega_{g}$, where $\omega_{g}\approx\frac{1}{3}\omega_{r}$, as shown in Fig.~\ref{fig:phase_space}(a). Neglecting noise, the dynamics of this system are governed by the Hamiltonian
\begin{eqnarray}
\hat{H}_\mathrm{lab}(t) &=& \frac{\hbar \omega_{0}}{2}\hat{S}_{z} + \hbar\omega_r\hat{a}^\dagger\hat{a}+ 2\hbar\Omega_{g}\cos(\omega_{g}t)\hat{S}_{z}\Big\{\hat{a} + \hat{a}^{\dagger} \Big\} \nonumber \\
&& +~2\hbar\Omega_{\mu}\hat{S}_{x}\Big\{\cos([\omega_{0} + \delta]t) + \cos([\omega_{0} - \delta]t) \Big\}, \nonumber \\
\end{eqnarray}
\noindent where $\Omega_{g}$ is the magnetic field gradient Rabi frequency and $\Omega_{\mu}$ is the microwave Rabi frequency. We define two-ion Pauli spin operators as $\hat{S}_{\gamma} \equiv \hat{\sigma}_{\gamma,1} + \hat{\sigma}_{\gamma,2}$, with $\gamma\in\{x,y,z,+,-\}$ and $z$ as the qubit quantization axis. The shared ion motion has creation and annihilation operators $\hat{a}^{\dagger}$ and $\hat{a}$.  This Hamiltonian can be generalized to multiple pairs of microwave tones or arbitrary gradient time dependence without significant complication (see Appendix \ref{sec:hamiltonian_appendix}). Transforming into the interaction frame with respect to $\hat{H}_0 = \hbar\omega_0\hat{S}_z/2 + \hbar\omega_r\hat{a}^\dagger\hat{a}$ and ignoring terms that oscillate with frequencies near $2\omega_{0}$ gives:
\begin{eqnarray}\label{eq:main_ham}
\hat{H}(t) &=& 2\hbar\Omega_{\mu}\hat{S}_{+}\cos{(\delta t)} + 2\hbar\Omega_{g}\cos(\omega_{g}t)\hat{S}_{z}\hat{a}e^{-i\omega_{r}t} + H.c.,  \nonumber \\
\end{eqnarray}
The oscillating gradient at $\omega_{g}$, in combination with the detuned microwaves, can give rise to two spin-motion-coupling sideband interactions, occurring when $\delta=\omega_{r}\pm\omega_{g}$, respectively \cite{srinivas_2018}. The proposed gate relies on this feature of the interaction, along with the fact that the bichromatic microwave pair combines to give an effective modulation of $\Omega_{\mu}$.  This scheme requires a magnetic-field-sensitive qubit transition instead of the  magnetic-field-insensitive ``clock'' transitions typically preferred for their long coherence times.  We envision storing quantum information in a clock qubit and transferring the state populations to a field-sensitive qubit only during times when an entangling gate is being carried out.  Alternatively, microwave fields can be applied to field-sensitive transitions to create dressed-state clock qubits \cite{timoney_2011}.

Typical experimental values of the amplitude $\Omega_{g}/2\pi$ are in the few kHz regime, whereas $\Omega_{\mu}/2\pi$ can be in the MHz regime  \cite{srinivas_2018}. It is therefore convenient to analyze the system dynamics in a frame that eliminates the large size disparity between the terms.  This can be done by transforming into the interaction picture with respect to the bichromatic microwaves \cite{jonathan_2000,roos_2008,sutherland_2019}:
\begin{eqnarray}\label{eq:h_int}
\hat{H}_{I}(t) = \hat{U}^{\dagger}(t)\hat{H}(t)\hat{U}(t) + i\hbar\dot{\hat{U}}^{\dagger}(t)\hat{U}(t),
\end{eqnarray}
where the frame transformation $\hat{U}(t)$, given by 
\begin{eqnarray}\label{eq:unitary}
\hat{U}(t) = \exp\Big\{-i\hat{S}_{x}\frac{\Omega_{\mu}\sin(\delta t)}{\delta} \Big\},
\end{eqnarray}
encapsulates the dynamics of the microwave pair (see appendix).  In the time-dependent transformed basis of this ``bichromatic'' interaction picture, the system dynamics between times $t_i$ and $t_f$, given by the propagator $\hat{T}_I(t_i,t_f)$ of $\hat{H}_I(t)$, appear simplified.  Crucially, however, if the microwave pair (as parameterized by $\Omega_\mu$) can be turned on and off in such a manner that $\hat{U}(t_i)=\hat{I}$ and $\hat{U}(t_f)=\hat{I}$, where $\hat{I}$ is the identity operator, then the state evolution given by $\hat{T}_I(t_i,t_f)$ applies to the lab frame basis as well as the transformed basis in the bichromatic interaction picture.  This condition on $\hat{U}(t_i)$ and $\hat{U}(t_f)$ can be achieved by ramping the microwave pair on and off slowly with respect to $1/\delta$, or by choosing the gate duration $t_{G}$ (during which the microwave pair is on) such that $t_{G}\delta$ is an integer multiple of $2\pi$.  Either method can be used, with realistic parameters, such that the fidelity of the final state in the lab frame basis and the interaction frame basis differ by less than $10^{-4}$. While the first method would be used in experiments, we use the second method in this paper (unless otherwise specified) because it is simpler for the numerical simulations. Making the above transformation on Eq.~(\ref{eq:main_ham}) gives \cite{sutherland_2019}: 
\begin{eqnarray}\label{eq:single_tone}
\hat{H}_{I}(t) &=& 2\hbar \Omega_{g}\cos(\omega_{g}t)\Big\{\hat{a} e^{-i\omega_{r}t} + \hat{a}^{\dagger}e^{i\omega_{r}t} \Big\}\Big\{ \hat{S}_{z}\Big[J_{0}\Big(\frac{4\Omega_{\mu}}{\delta}\Big)  \nonumber \\ 
&&+~2\sum_{n=1}^{\infty}J_{2n}\Big( \frac{4\Omega_{\mu}}{\delta}\Big)\cos(2n\delta t) \Big] \nonumber \\
&& +~2\hat{S}_{y}\sum^{\infty}_{n=1}J_{2n-1}\Big(\frac{4\Omega_{\mu}}{\delta}\Big)\sin([2n-1]\delta t) \Big\}.
\end{eqnarray}

\noindent We also note that, for simplicity, we have assumed all single qubit operations implemented in this work are instantaneous and error-free.

\section{Polychromatic gates}\label{sec:poly_1}

In this section, we discuss a protocol for tuning Eq.~(\ref{eq:single_tone}) to represent an entangling gate that is insensitive to motional and qubit decoherence. By setting the conditions
\begin{eqnarray}\label{eq:two_un}
4\delta &=& (\omega_{r} - \omega_{g}) - j\Delta, \nonumber \\
8\delta &=& (\omega_{r} + \omega_{g}) - (j+1)\Delta,
\end{eqnarray}
where $j$ is an integer, and $\Delta=2\pi/t_{G}$ is on the order of $\Omega_{g}$, the terms $\propto J_{4}$ and $\propto J_{8}$ become slowly varying in time with respect to all other terms in the sum. They will thus make the dominant contribution to the system dynamics, while the other terms that appear in Eq.~(\ref{eq:single_tone}) are significantly off-resonant, with contributions scaling as $(\Omega_{g}/\delta)^{2}$, where we note that $\Omega_{g}\ll\delta$. In the laboratory frame, the conditions in Eq.~(\ref{eq:two_un}) are equivalent to setting $\delta$ and $\omega_g$ to predominantly drive both the $\omega_{r}-\omega_{g}$ and $\omega_{r}+\omega_{g}$ sidebands simultaneously, as shown in Fig.~\ref{fig:phase_space}(c). 

Keeping only these near-resonant terms gives:
\begin{eqnarray}\label{eq:fancy_two_tone}
\hat{H}_{I}(t)&\simeq& \hbar\Omega_{g}\hat{S}_{z}\Big\{J_{4}\Big(\frac{4\Omega_{\mu}}{\delta}\Big)\Big(\hat{a}e^{-ij\Delta t} + \hat{a}^{\dagger}e^{ij\Delta t} \Big) \nonumber + \\ && J_{8}\Big(\frac{4\Omega_{\mu}}{\delta}\Big)\Big(\hat{a}e^{-i(j+1)\Delta t} + \hat{a}^{\dagger}e^{i(j+1)\Delta t} \Big) \Big\},
\end{eqnarray}
which resembles the form of the motionally robust polychromatic gates discussed in Refs.~\cite{haddadfarshi_2016,shapira_2018,webb_2018}. This will generate a gate with $K$ loops if we choose
\begin{eqnarray}\label{eq:big_delta}
\Delta = 4\Omega_{g}K^{1/2}\Big\{\frac{[J_{4}(\frac{4\Omega_{\mu}}{\delta})]^{2}}{j} + \frac{[J_{8}(\frac{4\Omega_{\mu}}{\delta})]^{2}}{j+1}\Big\}^{1/2}.
\end{eqnarray}
Since $J_{4}(4\Omega_{\mu}/\delta)$ and $J_{8}(4\Omega_{\mu}/\delta)$ are independent functions, we may optimize the amplitudes of their effective tones by setting the value of $\Omega_{\mu}/\delta$. For example, when $j=1$, we can engineer a gate that is robust to gate duration errors when $J_{8}(4\Omega_{\mu}/\delta)/J_{4}(4\Omega_{\mu}/\delta) = -1$ \cite{shapira_2018}, or the motionally robust gates of Refs.~\cite{haddadfarshi_2016,webb_2018} when $J_{8}(4\Omega_{\mu}/\delta)/J_{4}(4\Omega_{\mu}/\delta) = -2$ \cite{haddadfarshi_2016}. These gates are, unfortunately, still sensitive to time-varying qubit frequency shifts. 

\section{Qubit Decoherence}

We model the effects of time-dependent qubit frequency shifts by adding a term $\hat{H}_{z}(t) \equiv (\hbar\varepsilon/2) \cos(\omega_{\varepsilon}t)\hat{S}^{(\prime)}_{z}$ to the ideal Hamiltonian in Eq.~(\ref{eq:main_ham}). Here, we introduce an antisymmetric two-qubit Pauli operator $\hat{S}^{\prime}_{z} \equiv \hat{\sigma}_{z,1}-\hat{\sigma}_{z,2}$ in order to represent a differential qubit shift. In the bichromatic interaction picture, keeping only near-resonant terms (assuming $\varepsilon, \omega_\varepsilon\ll\delta$), $\hat{H}_{z,I}(t)$ can be written as:
\begin{eqnarray}\label{eq:qubit_shift}
\hat{H}_{z,I}(t) \simeq \frac{\hbar\varepsilon}{2}J_{0}\Big(\frac{4\Omega_{\mu}}{\delta} \Big)\cos(\omega_{\varepsilon}t)\hat{S}^{(\prime)}_{z}.
\end{eqnarray}
Note that the above equation is valid for symmetric and antisymmetric Pauli operators. When the value of $4\Omega_{\mu}/\delta$ is set to one of the zeros of the $J_{0}$ Bessel function, the expression in Eq.~(\ref{eq:qubit_shift}) goes to zero and the qubit frequency shifts do not contribute to the dynamics.  The off-resonant terms dropped from Eq.~(\ref{eq:qubit_shift}) all oscillate at integer multiples of $\delta$, such that in the typical case where $\varepsilon, \omega_\varepsilon\ll\delta$, their effect averages to zero; we refer to this phenomenon as intrinsic dynamical decoupling (IDD) \cite{sutherland_2019}. By tuning $\delta$, $\omega_{g}$, and $\omega_{r}$ so that Eqs.~(\ref{eq:two_un}) are met for a particular value of $j$, and setting $4\Omega_{\mu}/\delta \simeq 8.65$, the third IDD point (i.e. the third zero of $J_{0}(4\Omega_{\mu}/\delta)$), we perform a gate such that $J_{8}(4\Omega_{\mu}{\delta})/J_{4}(4\Omega_{\mu}/\delta) \simeq -1.22$ (see Fig.~\ref{fig:phase_space}(c)); we will refer to this as the IDD-$j$ gate. The phase space trajectory of an IDD-2 gate is shown in Fig.~\ref{fig:phase_space}(b). Figure~\ref{fig:phase_space}(b) also compares the IDD-2 gate's trajectory to that of a single-tone $\hat{\sigma}_{z}\hat{\sigma}_{z}$ gate corresponding to the $J_{2}$ resonance performed at the first IDD point (here referred to as IDD-single) \cite{sutherland_2019}. While the phase space trajectories of IDD-single and IDD-$1$ gates are not completely centered on the origin, those of IDD-$j$ gates for $j\geq2$ are, resulting in less time-averaged spin-motion entanglement, and consequently less impact of motional decoherence on gate fidelity. For all the calculations shown, we use experimentally achievable values of $\Omega_{g}/2\pi = 1$ kHz and $\omega_{r}/2\pi = 6.5$ MHz. Here $\omega_{g}/2\pi = 5$ MHz for the IDD-single gate; for the IDD-$j$ gates, $\omega_{g}$ is determined by solving Eqs.~(\ref{eq:two_un}) and (\ref{eq:big_delta}), giving $\omega_{g} \sim \omega_{r}/3$.

\begin{figure}[tb]
\includegraphics[width=0.48\textwidth]{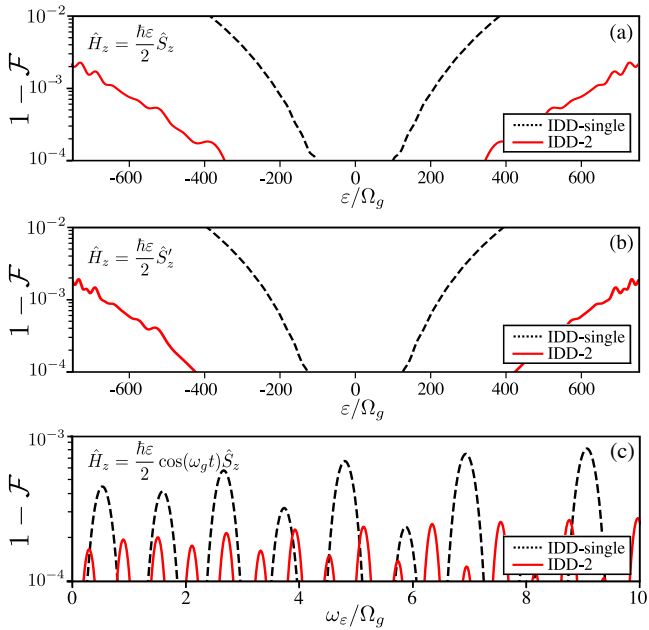}
\centering
\caption{(a) Gate infidelity $1- \mathcal{F}$ (plotted logarithmically) versus symmetric qubit frequency offset $\varepsilon$ for IDD-single (black dashed) and IDD-2 (solid red) gates with a given gradient strength $\Omega_{g}$. (b) Gate infidelity $1- \mathcal{F}$ versus antisymmetric qubit frequency offset $\varepsilon$ for IDD-single (black dashed) and IDD-2 (solid red) gates with a given gradient strength $\Omega_{g}$. (c) Gate infidelity $1-\mathcal{F}$ for the same gates versus the frequency $\omega_{\varepsilon}$ of a symmetric qubit frequency shift, with $\varepsilon=\Omega_{g}/5$. Every simulated gate has two phase space loops with a global qubit $\pi$ rotation in between.}
\label{fig:qubit_shift}
\end{figure}

In addition to the intrinsic dynamical decoupling effect described above, the IDD-$j$ gates can also be made highly insensitive to static qubit frequency shifts.  These gates produce an effective interaction of the form $\sigma_{z,1}\sigma_{z,2}$, such that qubit frequency shifts as shown in Eq.~(\ref{eq:qubit_shift}) commute with the gate operation.  As a result, the effect of static shifts can be removed (when $\varepsilon \ll \delta$) by performing a $K=2$ loop gate with a qubit $\pi$ rotation in between loops (a Walsh modulation of index 1) \cite{hayes_2012}. Figures~\ref{fig:qubit_shift}(a) and \ref{fig:qubit_shift}(b) show numerical simulations (including all terms in Eq.~(\ref{eq:main_ham})) of the impact of symmetric ($\propto \hat{S}_{z}$) and antisymmetric ($\propto \hat{S}^{\prime}_{z}$) static shifts, respectively, on the fidelity $\mathcal{F}$. Both of these show that the fidelity of our gates is unaffected by static frequency shifts at the the $10^{-4}$ level for $\varepsilon \lesssim 300~\Omega_{g}$. Note that the insensitivity to antisymmetric qubit frequency shifts implies that individual qubit addressing in frequency space should be possible with this scheme. We define the fidelity as $\mathcal{F}\equiv \bra{\Phi^{+}}\hat{\rho}(t_{G})\ket{\Phi^{+}}$, where $\ket{\Phi^{+}}\equiv1/\sqrt{2}\left(\ket{\downarrow\downarrow} + i\ket{\uparrow\uparrow}\right)$ is the target qubit state, when starting in $\ket{\downarrow\downarrow}$ and performing global $\pi/2$ rotations (perpendicular to $\hat{z}$) immediately before and after the $\sigma_{z,1}\sigma_{z,2}$ gate. This technique breaks down as $\varepsilon$ approaches $\delta$, such that the off-resonant terms that were dropped in Eq.~(\ref{eq:qubit_shift}) become non-negligible. Figure~\ref{fig:qubit_shift} also makes a comparison to the IDD-single gate proposed in Ref.~\cite{sutherland_2019}. Sensitivity to oscillating qubit frequency shifts is shown in Fig.~\ref{fig:qubit_shift}(c), where we plot the infidelity $1-\mathcal{F}$ versus $\omega_{\varepsilon}$ assuming a shift amplitude $\varepsilon = \Omega_{g}/5$. For $1-\mathcal{F}\ll1$, the infidelity scales as $\varepsilon^{2}$. The numerical simulations shown in Fig.~\ref{fig:qubit_shift} use a 20 $\mu$s Blackman envelope to shape the rising and falling edges of the gradient pulses and the microwave pulses \cite{blackman_1958}. We turn on the microwave pair first, ramping up the gradient after the microwaves reach steady state, and ramp down in the reverse order at the end of the pulses.  This is done because the gate speed is linear in the gradient strength, while it depends on the microwave amplitude as the argument of the two Bessel functions $J_{4}$ and $J_{8}$ (see Eq.~(\ref{eq:fancy_two_tone})), causing complicated undesired dynamics during the microwave ramp if the gradient is already present. The errors seen in Fig.~\ref{fig:qubit_shift}(c) occur during the microwave rise and fall times, when $\Omega_{\mu}$ is not at the IDD point, such that the qubits are vulnerable to frequency fluctuations of the form in $\hat{H}_{z}(t)$.  We note that qubit shifts that oscillate at or near $n\delta$ (for integer $n$) appear in the bichromatic interaction picture as static error terms $\propto \hat{S}_{z}$ (for even $n$) or $\propto \hat{S}_{y}$ (for odd $n$) \cite{sutherland_2019}.  Experimentally, qubit frequency fluctuations near $n\delta$ can arise from residual magnetic fields at $\omega_g$ from the currents generating the gradient. Choosing the $J_{4}$ and $J_{8}$ resonances to implement the gate makes $n$ even, and so the resulting errors are $\propto \hat{S}_{z}$ and can be removed as described above (see Sec.~\ref{sec:res_field}).

\begin{figure}[tb]
\includegraphics[width=0.495\textwidth]{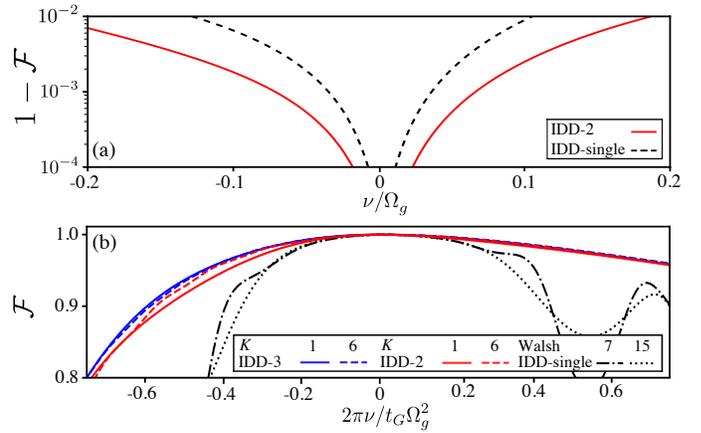}
\centering
\caption{(a) Gate infidelity $1-\mathcal{F}$ (plotted logarithmically) versus motional frequency offset $\nu$ for both of the qubit-frequency-shift-resistant gates shown in Figure~\ref{fig:qubit_shift}, given a fixed gradient $\Omega_{g}$. Both gates shown here undergo two phase space loops with a global qubit $\pi$ rotation in between. (b) Comparison of IDD-single gates performed with Walsh modulation of index 7 or 15 ($K=8$ and $K=16$, respectively) to IDD-2 and IDD-3 gates with $K=1$ or $K=6$, plotted against a dimensionless motional offset $\nu/\Omega_{g}$ normalized by a dimensionless gate duration $t_{G}\Omega_{g}/2\pi$. All IDD gates fall on a similar curve, as do IDD-single gates for small values of $2\pi\nu/t_{G}\Omega_{g}^{2}$.}
\label{fig:mot_shift}
\end{figure}

\section{Motional decoherence}

Gates with multiple blue and red sideband pairs, such as those presented in this work, have reduced sensitivity to motional frequency offsets. For example, an IDD-$j$ gate is a linear superposition of a $j$ and a $j$+1 loop IDD-single gate, each with amplitudes of opposite signs. A motional frequency offset $\nu$ shifts the secular frequency such that $\omega_{r}\rightarrow \omega_{r} + \nu$, resulting in a residual displacement in phase space at the end of the gate. With nonzero $\nu$, the superposed gates experience opposite displacements in phase space which coherently cancel each other. As shown in Fig.~\ref{fig:mot_shift}(a), this results in reduced sensitivity to $\nu$ \cite{webb_2018,shapira_2018}.  Fig.~\ref{fig:mot_shift}(b) shows that this coherent error cancellation can provide reduced sensitivity to $\nu$ by increasing either the number of loops $K$ or the order $j$ of the gate. The infidelity due to an offset $\nu$ will remain constant for an increased $\nu$ if $t_{G}$ is also increased proportionally; this remains true whether the increased $t_{G}$ is associated with more loops $K$ or larger values of $j$. In Figure~\ref{fig:mot_shift}(b), we plot Bell state infidelity for a variety of different gates versus the dimensionless motional frequency offset $\nu/\Omega_{g}$, normalized by the dimensionless gate duration $t_{G}\Omega_{g}/2\pi$.  With this normalization, the different gates fall on approximately the same curve of sensitivity to motional frequency offsets. As the gate-time-normalized motional frequency offset becomes larger, the IDD-$j$ gates have higher fidelity than single tone gates following Walsh sequences, shown in Fig.~\ref{fig:mot_shift}(b) \cite{hayes_2012}.

\begin{figure}[tb]
\includegraphics[width=0.45\textwidth]{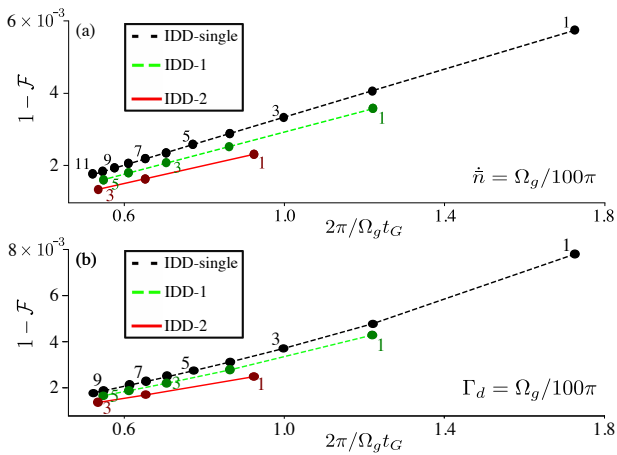}
\centering
\caption{(a) Infidelity $1-\mathcal{F}$ due to motional heating versus normalized gate speed $2\pi/\Omega_{g}t_{G}$. Point labels indicate the number of phase space loops $K$.  For a given gradient strength $\Omega_{g}$ and heating rate $\dot{\bar{n}} = \Omega_{g}/100\pi$, the error of a gate scales with $1/t_{G}$. For a given $t_{G}$, the IDD-2 (solid red line) and IDD-1 (dashed green line) gates outperform the IDD-single (dashed black line) gate.  Higher-order IDD-$j$ gates for $j\geq3$ are similar to the IDD-2 gate. Panel (b) shows the same calculations for motional dephasing instead of motional heating, assuming a motional dephasing rate of $\Gamma_{d} = \Omega_{g}/100\pi$.}
\label{fig:heating}
\end{figure}

In Fig.~\ref{fig:heating}, we show the increased robustness to heating and motional dephasing of the IDD-$j$ gates. We treat these decoherence mechanisms as Markovian, using a Lindblad formalism \cite{breuer_2002}. Note that for these calculations, we use Eq.~(\ref{eq:fancy_two_tone}) to calculate the infidelity; this gives the same motional decoherence effect as the full integration of Eq.~(\ref{eq:main_ham}) (see appendix). Geometric phase gates can be made less sensitive to motional heating by performing more phase space loops, with $1-\mathcal{F}$ scaling $\propto 1/t_{G}$. Therefore, in order to make a relevant comparison to our polychromatic gates, Fig.~\ref{fig:heating}(a) shows the infidelity due to a heating rate of $\dot{\bar{n}}= \Omega_{g}/100\pi$, versus $2\pi/\Omega_{g}t_{G}$, for the IDD-single, IDD-1, and IDD-2 gates. Similarly, in Fig.~\ref{fig:heating}(b) we compare the same set of gates for a motional dephasing rate of $\Gamma_{g} = \Omega_{g}/100\pi$. Both calculations in Fig.~\ref{fig:heating} show that, while better than the IDD-single, the IDD-1 gate is not as robust as the IDD-2; this can be understood because the IDD-1 trajectory is not centered on the origin of phase space. For $j>2$, however, we find that there is not a significant improvement of $\mathcal{F}$ versus $t_{G}$ relative to $j=2$. This is because the phase space trajectories of the IDD-$j$ gates with $j>1$ are all centered on the origin, thus saturating improvement to the time-averaged spin-motion entanglement.

\section{Residual magnetic field from gradient}\label{sec:res_field}
Generating a gradient that oscillates at $\omega_{g}$ typically leads to a residual magnetic field that also oscillates at $\omega_{g}$. We represent the Hamiltonian of this residual magnetic field as
\begin{eqnarray}
\hat{H}_{z}(t) = 2\hbar\Omega_{z}\cos(\omega_{g}t)\hat{S}_{z},
\end{eqnarray}
which adds to the full Hamiltonian, given by Eq.~(\ref{eq:main_ham}). To analyze this field's effect on our gate, we make the transformation given by Eq.~(\ref{eq:unitary}). This gives:
\begin{eqnarray}\label{eq:shaun4ever}
\hat{H}_{z,I}(t) &= & 2\hbar\Omega_{z}\cos(\omega_{g}t)\times \nonumber \\
&& \Big\{ \hat{S}_{z}\Big[J_{0}\Big(\frac{4\Omega_{\mu}}{\delta}\Big) + 2\sum_{n=1}^{\infty}J_{2n}\Big( \frac{4\Omega_{\mu}}{\delta}\Big)\cos(2n\delta t) \Big] \nonumber \\
&& +~2\hat{S}_{y}\sum^{\infty}_{n=1}J_{2n-1}\Big(\frac{4\Omega_{\mu}}{\delta}\Big)\sin([2n-1]\delta t) \Big\}.\nonumber \\
\end{eqnarray}
Assuming $\Omega_{z} \ll \delta$, we can simplify this equation by dropping far-off-resonant terms. By solving Eq.~(\ref{eq:two_un}), we find that:
\begin{eqnarray}
\delta = \frac{\omega_{g}}{2} - \frac{\Delta}{4}.
\end{eqnarray}
Keeping only the near-resonant terms gives:
\begin{eqnarray}
\hat{H}_{I,z}(t) \simeq 2\hbar\Omega_{z}J_{2}\Big(\frac{4\Omega_{\mu}}{\delta} \Big)\cos(\Delta t/2)\hat{S}_{z}.
\end{eqnarray}
Since the residual term is $\propto \hat{S}_{z}$, it commutes with the gate and its effect at the end of a single-loop gate, $t_{G}=2\pi/\Delta$, can be described by a unitary operator given by:
\begin{eqnarray}
\Hat{U}_{z}(t) &=& \exp\Big( -\frac{i}{\hbar}\int^{t_{G}}_{0}dt^{\prime}H_{I,z}(t^{\prime})\Big) \nonumber \\
&=& \exp\Big(-\frac{4i\Omega_{z}}{\Delta}J_{2}\Big(\frac{4\Omega_{\mu}}{\delta}\Big)\sin(\pi)\hat{S}_{z} \Big) \nonumber \\
&=& \hat{I}.
\end{eqnarray}
Thus, a residual homogenous magnetic field at $\omega_{g}$ (such that $\Omega_{z}\ll\delta$) does not impact the fidelity of the gate.  For the gates described in this work, the $J_{4}$ and $J_{8}$ resonances were chosen specifically because the near-resonant ($\sim \omega_{g}$) contributions are $\propto \hat{S}_{z}$ (see Eq.~(\ref{eq:shaun4ever})). One could imagine a similar scheme using the $\omega_{r}\pm\omega_{g}$ sidebands, and the $J_{2}$ and $J_{4}$ resonances. This, however, results in an on-resonant term $\propto\hat{S}_{x}$ in $H_{I,z}$, as opposed to the term $\propto \hat{S}_{z}$ in the $J_{4}$, $J_{8}$ gate. Since this term does not commute with the gate, it would have to be nulled in order to not affect the fidelity, which would make the scheme significantly more difficult to implement.

\section{Polychromatic gates with multiple microwave pairs}\label{sec:poly_2}

The gate presented in this work shows similar resilience to motional decoherence as the laser-free polychromatic gate successfully demonstrated by Webb et al. in Ref.~\cite{webb_2018}, but it also has important advantages. In their experiment, Webb et al. implemented laser-free polychromatic gates by tuning two bichromatic microwave pairs close to the frequency of the sideband of a static gradient ($\omega_{g} = 0$).  This scheme requires a total of 12 oscillating control fields plus a strong static magnetic field gradient \cite{webb_2018}, whereas the scheme presented here requires only two oscillating control fields, plus a strong, oscillating magnetic field gradient. Importantly, the scheme of Ref.~\cite{webb_2018} produces unavoidable errors outside the $\Omega_{\mu}/\omega_{r} \ll 1$ regime, which limits the achievable gate speed, while our gate scheme has no such limitation.  We derive this result using our interaction picture formalism described in the appendix.  

\begin{figure}[h]
\includegraphics[width=0.5 \textwidth]{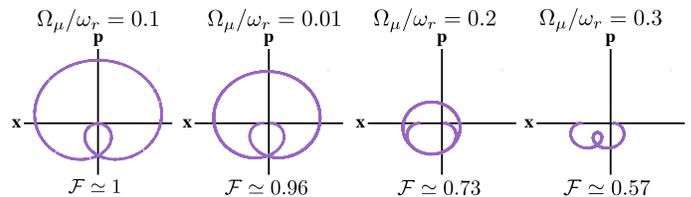}
\centering
\caption{Phase space trajectories and fidelities $\mathcal{F}$, calculated using Eq.~(\ref{eq:main_ham}), for increasing values of $\Omega_{\mu}/\omega_{r}$, where $\Omega_{\mu}$ is the microwave strength and $\omega_{r}$ is the motional frequency. Even without imperfections, when $\Omega_{\mu}\sim\omega_{r}$ the phase space trajectory distorts and the value of $\mathcal{F}$ decreases.}
\label{fig:sussex_phase_space}
\end{figure}

Outside of this regime, terms that are higher-order in $\Omega_{\mu}/\omega_{r}$, as seen in Eq.~(\ref{eq:compli_pauli}), contribute to the dynamics and distort the phase space trajectory. In particular, the appearance of terms that oscillate at exactly $\omega_{r}$ prevent the motion from returning to its original state, leaving residual entanglement. This suggests that the implementation in Ref.~\cite{webb_2018} is limited to the weak microwave field regime. We illustrate this effect in Fig.~\ref{fig:sussex_phase_space}, where we simulate the trajectory of Eq.~(\ref{eq:compli_pauli}) for a static gradient with increasing values of $\Omega_{\mu}$, with no added noise. We can see that for the $\Omega_{\mu}/\omega_{r}\ll 1$ regime, the phase space trajectory follows the two-tone trajectory described in Ref.~\cite{haddadfarshi_2016} and $\mathcal{F} \rightarrow 1$. When $\Omega_{\mu}/\omega_{r} \sim 1$, however, the trajectory becomes distorted and the gate fidelity decreases. Since the speed of the gate in the $\Omega_{\mu} \ll \delta$ regime is $\propto \Omega_{\mu}$, this indicates that for a limited value of $\Omega_{g}$\textemdash typically the case in laser-free experiments\textemdash the maximum speed of a polychromatic gate performed via multiple pairs of microwave fields is slower than can be achieved with the oscillating gradient method presented here.

\section{Conclusion}

In this work, we described a new type of laser-free trapped ion entangling gate that can be tuned to be simultaneously robust to motional decoherence and qubit frequency shifts. The gate requires only two microwave magnetic fields and one near-motional-frequency magnetic field gradient. In addition to using analytic expressions, we numerically demonstrated this robustness. The design should enable higher gate fidelities for laser-free entangling gates without substantially increasing gate duration or the complexity of the required driving fields. We further demonstrated the advantages of our gate by showing that generating laser-free, polychromatic entangling gates using multiple pairs of microwave fields grows increasingly difficult when the microwave Rabi frequencies become comparable to the trap's secular frequency.

\section*{Acknowledgments}

We acknowledge helpful discussions with Y. Rosen, D.M. Lucas, and S.C. Webster, and thank D.C. Cole and K. Cui for careful reading of the manuscript.  R.S., S.C.B., H.M.K., and D.T.C.A. are Associates in the Professional Research Experience Program (PREP) operated jointly by NIST and the University of Colorado Boulder under award 70NANB18H006 from the U.S. Department of Commerce, National Institute of Standards and Technology.  This work was supported by ARO, ONR, and the NIST Quantum Information Program.  This paper is a partial contribution of NIST and is not subject to US copyright. Part of this work was performed under the auspices of the U.S. Department of Energy by Lawrence Livermore National Laboratory under Contract DE-AC52-07NA27344. LLNL-JRNL-792460

\appendix

\section{\label{sec:hamiltonian_appendix}Laser-free Hamiltonian}
We consider a general Hamiltonian for laser-free gates between two trapped ions with identical qubit frequencies:
\begin{eqnarray}\label{eq:labframe}
\hat{H}_\mathrm{lab}(t) &=& \frac{\hbar \omega_{0}}{2}\hat{S}_{z} + \hbar\omega_r\hat{a}^\dagger\hat{a}+ 2\hbar\Omega_{g}f(t)\hat{S}_{j}\Big\{\hat{a} + \hat{a}^{\dagger} \Big\} \nonumber \\
&& +~2\hbar\Omega_{\mu}\hat{S}_{i}\sum_{n}c_{n}\Big\{\cos([\omega_{0} + \delta_{n}]t) \nonumber \\
&& +~\cos([\omega_{0} - \delta_{n}]t) \Big\}, \nonumber \\
\end{eqnarray}
\noindent where $i \in \{x,y\}$ and $j\in\{x,y,z\}$. We define two-ion Pauli spin operators as $\hat{S}_{\gamma} \equiv \hat{\sigma}_{\gamma,1} + \hat{\sigma}_{\gamma,2}$. We have taken $z$ to be the qubit quantization axis and $\omega_0$ to be the qubit frequency. Here we have assumed an ion crystal whose internal states are coupled via a motional mode with frequency $\omega_{r}$ and creation (annihilation) operators $\hat{a}^{\dagger}\,(\hat{a})$.  As they appear in this equation, $\Omega_{\mu}$ and $\Omega_{g}$ are Rabi frequencies. Here, $c_{n}$ is the amplitude (order unity) of the $n^{th}$ pair of equal amplitude fields, symmetrically detuned from the qubit frequency by $\pm\,\delta_{n}$. The $\Omega_{g}$ term describes the gradient-induced coupling of the internal states to the motion. The time dependent function of the gradient, $f(t)$, can be arbitrary; here, we take it to be constant or sinusoidally oscillating with a frequency of the same order of magnitude as $\omega_{r}$.

We transform Eq.~(\ref{eq:labframe}) into the interaction picture with respect to the harmonic motion and qubit frequency terms, $\hat{H}_0 = \hbar\omega_0\hat{S}_z/2 + \hbar\omega_r\hat{a}^\dagger\hat{a}$. Furthermore, we make the rotating wave approximation to eliminate terms oscillating with frequencies near $2\omega_{0}$. This gives the interaction Hamiltonian:
\begin{eqnarray}\label{eq:app_rwa}
\hat{H}(t) &=& 2\hbar\Omega_{\mu}\hat{S}_{i}\sum_{n}c_{n}\cos(\delta_{n} t) \nonumber \\ 
&&+ 2\hbar\Omega_{g}f(t)\hat{S}_{j}\Big\{\hat{a}e^{-i\omega_{r}t} + \hat{a}^{\dagger}e^{i\omega_{r}t} \Big\},
\end{eqnarray}
which is Eq.~(\ref{eq:main_ham}) from the main text when $i=x$ and $j=z$. Note that when $j\in\{x,y\}$, $f(t)$ changes from Eq.~(\ref{eq:labframe}) to Eq.~(\ref{eq:app_rwa}) when frequencies near $2\omega_{0}$ are dropped (see Ref.~\cite{sutherland_2019} for more details).

\section{Bichromatic interaction picture}

For completeness, we summarize an analysis of geometric phase gates for large microwave magnetic fields ($\Omega_{\mu}\sim\omega_{r}$) demonstrated originally in Ref.~\cite{sutherland_2019}. Assuming a Hamiltonian that takes the form of Eq.~(\ref{eq:main_ham}), representing a system with a bichromatic magnetic field pair and a (static or oscillatory) magnetic field gradient, we obtain:
\begin{eqnarray}
\hat{H}(t) &=& \hat{H}_{\mu}(t) + \hat{H}_{g}(t) \nonumber \\
&=& 2\hbar\Omega_{\mu}\hat{S}_{i}\cos(\delta t) + 2\hbar\Omega_{g}f(t)\hat{S}_{j}\Big\{\hat{a}e^{-i\omega_{r}t} + \hat{a}^{\dagger}e^{i\omega_{r}t} \Big\}, \nonumber \\
\end{eqnarray}
\noindent where we have dropped the subscript of $\delta$, since there is only one microwave pair. In order to exactly account for the dynamics caused by the larger of the two terms in this equation ($\hat{H}_{\mu}(t)$, the bichromatic field term), we note that the time propagator for this field alone is exactly solvable:
\begin{eqnarray}
\hat{U}(t) &=& \exp\Big(-2i\Omega_{\mu}\hat{S}_{i}\int^t_{0}dt^{\prime}\cos(\delta t^{\prime})\Big) \nonumber \\
&=& \exp\Big(-\frac{2i\Omega_{\mu}}{\delta}\sin(\delta t)\hat{S}_{i}\Big). 
\end{eqnarray}

We can then transform into the interaction picture with respect to the bichromatic microwave term:

\begin{eqnarray}
\hat{H}_{I} &=& \hat{U}^{\dagger}(t)\hat{H}_{g}(t)\hat{U}(t) \nonumber \\
&=& 2\hbar\Omega_{g}f(t)\Big\{\hat{a}e^{-i\omega_{r}t} + \hat{a}^{\dagger}e^{i\omega_{r}t} \Big\}\hat{U}^{\dagger}(t)\hat{S}_{j}\hat{U}(t). \nonumber \\
\end{eqnarray}
\noindent Applying the Jacobi-Anger expansion (see Ref.~\cite{sutherland_2019}), 

\begin{eqnarray}\label{eq:rotate_sj}
\hat{U}^{\dagger}(t)\hat{S}_{j}\hat{U}(t) &=& \Big\{ \hat{S}_{j}\Big[J_{0}\Big(\frac{4\Omega_{\mu}}{\delta}\Big) + 2\sum_{n=1}^{\infty}J_{2n}\Big( \frac{4\Omega_{\mu}}{\delta}\Big)\cos(2n\delta t) \Big] \nonumber\\ 
&&-~2\epsilon_{ijk}\hat{S}_{k}\sum^{\infty}_{n=1}J_{2n-1}\Big(\frac{4\Omega_{\mu}}{\delta}\Big)\sin([2n-1]\delta t) \Big\}, \nonumber \\
\end{eqnarray}

\noindent giving

\begin{eqnarray}
\hat{H}_{I}(t) &=& 2\hbar \Omega_{g}f(t)\Big\{\hat{a} e^{-i\omega_{r}t} + \hat{a}^{\dagger}e^{i\omega_{r}t} \Big\}\Big\{ \hat{S}_{j}\Big[J_{0}\Big(\frac{4\Omega_{\mu}}{\delta}\Big) \nonumber \\ 
&& +~2\sum_{n=1}^{\infty}J_{2n}\Big( \frac{4\Omega_{\mu}}{\delta}\Big)\cos(2n\delta t) \Big] \nonumber
\\ && -~2\epsilon_{ijk}\hat{S}_{k}\sum^{\infty}_{n=1}J_{2n-1}\Big(\frac{4\Omega_{\mu}}{\delta}\Big)\sin([2n-1]\delta t) \Big\}. \nonumber \\
\end{eqnarray}
\noindent This shows a series of resonances corresponding to $\hat{\sigma}_{z}\hat{\sigma}_{z}$ and $\hat{\sigma}_{\varphi}\hat{\sigma}_{\varphi}$ (M\o lmer-S\o renson) gates when the sideband frequency is an even or odd integer multiple of $\delta$, respectively.

\section{Polychromatic gates with multiple pairs of microwaves}

The Hamiltonian for the system containing multiple microwaves pairs used in \cite{webb_2018} is given by:

\begin{eqnarray}\label{eq:two_tone_ham}
\hat{H}(t) &=& \hat{H}_{\mu,1}(t) + \hat{H}_{\mu,2} + \hat{H}_{g}(t) \nonumber \\
&=& 2\hbar\Omega_{\mu}\hat{S}_{x}c_{1}\cos(\delta_{1} t) + 2\hbar \Omega_{\mu}\hat{S}_{x}c_{2}\cos(\delta_{2} t) \nonumber \\
&& +~2\hbar\Omega_{g}\hat{S}_{z}\Big\{\hat{a}e^{-i\omega_{r}t} + \hat{a}^{\dagger}e^{i\omega_{r}t} \Big\}.
\end{eqnarray}
It is helpful to study the dynamics in the interaction picture with respect to both of the microwave field terms, which are exact. This gives:
\begin{eqnarray}\label{eq:double_int}
    H_{I}(t) = \hat{U}^{\dagger}_{1}(t)\hat{U}^{\dagger}_{2}(t)\hat{H}_{g}(t)\hat{U}_{2}(t)\hat{U}_{1}(t),
\end{eqnarray}
where
\begin{eqnarray}
U_{j}(t) &= \nonumber& \exp\Big(-\frac{i}{\hbar}\int^{t}_{0}dt^{\prime}H_{\mu,j}(t^{\prime}) \Big) \\
&=& \exp\Big(-\frac{2i\Omega_{\mu}c_{j}}{\delta_{j}}\sin(\delta_{j}t)\hat{S}_{x} \Big).
\end{eqnarray}

\noindent Focusing on the Pauli operators, and using Eq.~(\ref{eq:rotate_sj}) twice we obtain:
\begin{eqnarray}\label{eq:compli_pauli}
&& \hat{U}^{\dagger}_{1}(t)\hat{U}^{\dagger}_{2}(t)\hat{S}_{z}\hat{U}_{2}(t)\hat{U}_{1}(t)  = \nonumber \\
&& \hat{S}_{z}\Big\{\Big[J_{0}\Big(\frac{4\Omega_{\mu}c_{1}}{\delta_{1}}\Big) + 2\sum_{n=1}^{\infty}J_{2n}\Big(\frac{4\Omega_{\mu}c_{1}}{\delta_{1}}\Big)\cos(2n\delta_{1}t)\Big] \nonumber \\
&& ~\times \Big[ J_{0}\Big(\frac{4\Omega_{\mu}c_{2}}{\delta_{2}}\Big) + 2\sum_{n=1}^{\infty}J_{2n}\Big(\frac{4\Omega_{\mu}c_{2}}{\delta_{2}}\Big)\cos(2n\delta_{2}t) \Big] \nonumber \\
&& -~4\sum_{n,n^{\prime}=1}^{\infty}J_{2n-1}\Big(\frac{4\Omega_{\mu}c_{1}}{\delta_{1}} \Big)J_{2n^{\prime}-1}\Big(\frac{4\Omega_{\mu}c_{2}}{\delta_{2}} \Big) \nonumber \\ 
&& \times \sin([2n-1]\delta_{1} t) \sin([2n^{\prime}-1]\delta_{2}t) \Big\} \nonumber \\
&& +~2\hat{S}_{y}\Big\{\Big[\sum_{n=1}^{\infty}J_{2n-1}\Big(\frac{4\Omega_{\mu}c_{1}}{\delta_{1}} \Big)\sin([2n-1]\delta_{1}t)\Big] \nonumber \\
&& \times \Big[J_{0}\Big(\frac{4\Omega_{\mu}c_{2}}{\delta_{2}} \Big) + 2\sum_{n=1}^{\infty}J_{2n}\Big(\frac{4\Omega_{\mu}c_{2}}{\delta_{2}}\Big) \cos(2n\delta_{2}t)\Big] \nonumber \\
&& + \Big[\sum_{n=1}^{\infty}J_{2n-1}\Big(\frac{4\Omega_{\mu}c_{2}}{\delta_{2}} \Big)\sin([2n-1]\delta_{2}t)\Big] \nonumber \\
&& \times \Big[J_{0}\Big(\frac{4\Omega_{\mu}c_{1}}{\delta_{1}} \Big) + 2\sum_{n=1}^{\infty}J_{2n}\Big(\frac{4\Omega_{\mu}c_{1}}{\delta_{1}} \Big)\cos(2n\delta_{1}t) \Big] \Big\}. \nonumber \\ 
\end{eqnarray}
This equation can be simplified for the case considered by Webb et al. \cite{webb_2018}, where $\Omega_{\mu}\ll \omega_{r}$, $\delta_{1} = \omega_{r} - \Delta$, $\delta_{2} = \omega_{r} - 2\Delta$. Here $\Delta$ is of the same order as $\Omega_{g}$. Noting that $J_{n}(x) \rightarrow (\frac{x}{2})^{n}/n!$ as $x\rightarrow 0$, and keeping only the terms that are first order in $\Omega_{\mu}/\omega_{r}$, we obtain:
\begin{eqnarray}
&& \hat{U}^{\dagger}_{1}(t)\hat{U}^{\dagger}_{2}(t)\hat{S}_{z}\hat{U}_{2}(t)\hat{U}_{1}(t) \nonumber \\
&& \simeq \hat{S}_{z} + \frac{4\Omega_{\mu}}{\omega_{r}}\hat{S}_{y}\Big\{c_{1}\sin([\omega_{r} - \Delta]t) + c_{2}\sin([\omega_{r}-2\Delta]t)\Big\},  \nonumber \\
\end{eqnarray}
which, if we neglect the terms oscillating at frequencies on the order of $\omega_{r}$, we get an interaction picture Hamiltonian:
\begin{eqnarray}
\hat{H}_{I}(t) &=& \frac{4i\hbar\Omega_{\mu}\Omega_{g}}{\omega_{r}}\hat{S}_{y}\Big\{\hat{a}^{\dagger}(c_{1}e^{i\Delta t} + c_{2}e^{2i\Delta t}) \nonumber \\
&& -~\hat{a}(c_{1}e^{-i\Delta t} + c_{2}e^{-2i\Delta t}) \Big\},
\end{eqnarray}

\noindent giving the form of a polychromatic gate \cite{haddadfarshi_2016}. 
This means that in the $\Omega_{\mu} \ll \omega_{r}$ limit, as is the case in Ref.~\cite{webb_2018}, this is a valid technique for performing laser-free polychromatic gates. If terms in Eq.~(\ref{eq:compli_pauli}) that are higher-order in $\Omega_{\mu}/\omega_{r}$ contribute, terms that oscillate at exactly $\omega_{r}$ appear, such as 
\begin{eqnarray}
&& J_{1}\Big(\frac{4\Omega_{\mu}c_{2}}{\omega_{r}} \Big)J_{2}\Big(\frac{4\Omega_{\mu}c_{1}}{\omega_{r}} \Big)\sin([\omega_{r}-2\Delta]t)\cos(2[\omega_{r} - \Delta]t) \nonumber \\
&=& J_{1}\Big(\frac{4\Omega_{\mu}c_{2}}{\omega_{r}} \Big)J_{2}\Big(\frac{4\Omega_{\mu}c_{1}}{\omega_{r}} \Big)\Big\{ \sin([3\omega_{r} - 4\Delta]t)  - \sin(\omega_{r}t)\Big\}. \nonumber \\
\end{eqnarray}
These terms, as discussed in the text, prevent the trajectory from returning to the origin in phase space. \\

\section{Decoherence due to Bath Coupling}
\label{sec:bath}
In this section, we show that for the system described in the paper, we can calculate the infidelities due to Markovian heating and motional dephasing by numerically calculating the (significantly less computationally intensive) master equation for $\hat{H}_{I}$ (neglecting terms rotating near $\omega_{r}$) rather than the full Hamiltonian given by Eq.~(\ref{eq:main_ham}). This is because the unitary transformation given by Eq.~(\ref{eq:unitary}) commutes with the Lindblad operators that represent heating and motional dephasing. The full master equation for the time evolution of the full Hamiltonian, Eq.~(\ref{eq:main_ham}), including heating and motional dephasing, is:
\begin{eqnarray}\label{eq:master_equation}
\dot{\hat{\rho}}(t) &=& -\frac{i}{\hbar}[\hat{H}(t),\hat{\rho}(t)] + \dot{\bar{n}}\Big\{\hat{a}\hat{\rho}(t)\hat{a}^{\dagger} + \hat{a}^{\dagger}\hat{\rho}(t)\hat{a} \nonumber \\
&&- \frac{1}{2}(\hat{a}^{\dagger}\hat{a} + \hat{a}\hat{a}^{\dagger})\hat{\rho}(t) - \frac{1}{2}\hat{\rho}(t)(\hat{a}^{\dagger}\hat{a} + \hat{a}\hat{a}^{\dagger} )\Big\} \nonumber \\
&& + \Gamma_{d}\Big\{\hat{a}^{\dagger}\hat{a}\hat{\rho}(t)\hat{a}^{\dagger}\hat{a} - \frac{1}{2}(\hat{a}^{\dagger}\hat{a})^{2}\hat{\rho}(t) -\frac{1}{2}\hat{\rho}(t)(\hat{a}^{\dagger}\hat{a})^{2}\Big\}, \nonumber \\ 
\end{eqnarray}
where $\dot{\bar{n}}$ is the heating rate of the system and $\Gamma_{d}$ is the motional dephasing rate. Making the unitary transformation defined by Eq.~(\ref{eq:unitary}), we can solve for the density matrix in the interaction picture:
\begin{eqnarray}
\hat{\rho}_{I}(t) = \hat{U}^{\dagger}(t)\hat{\rho}(t)\hat{U}(t).
\end{eqnarray}
Since the above transformation commutes with both the heating and motional dephasing operators, the interaction picture master equation is simply:
\begin{eqnarray}\label{eq:mast_interact}
\dot{\hat{\rho}}_{I}(t) &=& -\frac{i}{\hbar}[\hat{H}_{I}(t),\hat{\rho}_{I}(t)] + \dot{\bar{n}}\Big\{\hat{a}\hat{\rho}_{I}(t)\hat{a}^{\dagger} + \hat{a}^{\dagger}\hat{\rho}_{I}(t)\hat{a} \nonumber \\
&&-~\frac{1}{2}(\hat{a}^{\dagger}\hat{a} + \hat{a}\hat{a}^{\dagger})\hat{\rho}_{I}(t) - \frac{1}{2}\hat{\rho}_{I}(t)(\hat{a}^{\dagger}\hat{a} + \hat{a}\hat{a}^{\dagger} )\Big\} \nonumber \\
&& +~\Gamma_{d}\Big\{\hat{a}^{\dagger}\hat{a}\hat{\rho}_{I}(t)\hat{a}^{\dagger}\hat{a} - \frac{1}{2}(\hat{a}^{\dagger}\hat{a})^{2}\hat{\rho}_{I}(t) -\frac{1}{2}\hat{\rho}_{I}(t)(\hat{a}^{\dagger}\hat{a})^{2}\Big\}. \nonumber \\ 
\end{eqnarray}
\noindent The fast rotating terms in $H_{I}$ can be dropped with an effect scaling $\propto (\Omega_{g}/\omega_{r})^{2}$. This is demonstrated by the calculations of the full numerical Hamiltonian shown for the pure state systems in this work. Numerically integrating Eq.~(\ref{eq:main_ham}) allows us to calculate $\hat{\rho}_{I}(t_{G})$, where $t_{G}$ is the gate duration, while accurately including the effects of Markovian heating and motional dephasing. As with the pure state calculations, when $\hat{U}(t_{G})\rightarrow \hat{I}$, $\hat{\rho}_{I}(t_{G}) \rightarrow \hat{\rho}(t_{G})$, meaning that our reduced calculation gives the same answer as would performing the full numerical integration of Eq.~(\ref{eq:master_equation}), as long as the pure state calculation is also valid.

\bibliography{biblio}

\end{document}